\documentstyle[preprint,aps,epsf]{revtex}

\tighten
\begin{document}
\draft

\title{Vortex Lines or Vortex-Line Chains at the Lower Critical Field in Anisotropic Superconductors?}     

\author{ W. A. M. Morgado\footnote{Corresponding author. Fax: 55 21
562-7368. e-mail: 
welles@if.ufrj.br}, M. M. Doria, and G. Carneiro}
\address{Instituto de F\'{\i}sica\\Universidade
Federal do Rio de Janeiro\\   C.P. 68528\\ 21945-970, Rio de
Janeiro-RJ, Brazil} 
\date{\today}
\maketitle

\begin{abstract}

The  vortex state at the lower critical field,
$H_{c1}$, in clean anisotropic superconductors  placed 
in an  external field tilted with respect to the axis of anisotropy
(c-axis)  is considered assuming two  possible arrangements: dilute
vortex-lines or dilute  vortex-line chains. 
By minimizing the Gibbs free energies in the London limit for each
possibility we obtain the corresponding lower critical fields as a
function of the tilt angle. The equilibrium configuration at $H_{c1}$
for a given tilt angle is identified with that for which $H_{c1}$ is
the smallest. We report results for parameter values typical of strong
and moderate anisotropy. We find that for strong anisotropy vortex-line
chains are favored for small tilt angles ($< 7.9^{\rm o}$) and
that at $7.9^{\rm o}$ there is coexistence between this configuration
and a vortex-line one. For moderate anisotropy we find that there is
little difference between the vortex-line and the vortex-chain lower
critical fields.

\end{abstract}

\narrowtext


Motivated by the discovery of  high-T$_c$ superconductivity in the
cuprates, calculations  of  vortex properties in uniaxially
anisotropic superconductors were carried out by several
workers \cite{rev1,rev2}. One surprising result is that the interaction
between a pair of straight vortex lines parallel to each other, tilted
and coplanar with respect to the c-axis is attractive at distances
larger than a value of the order of the penetration depth, as was first found by Grishin, Martynovich and  Yampol'skii\cite{gris,vatr1,vatr2,vatr3}. As a consequence, a  chain of such lines has
lower energy than the same vortex lines placed far apart. 
On the basis of this result it was suggested that at $H_{c1}$ the
mixed state consists of a vanishing small density of vortex-line chains
instead of  vortex lines \cite{vatr1,vatr3}. 

However, this is not correct in general for tilted vortex lines. The
reason is that in order to determine the $H_{c1}$
configuration it is necessary to  minimize   the
Gibbs free energy in the limit of vanishing vortex density for each
candidate vortex arrangement and, from it, obtain the corresponding $H_{c1}$.
The equilibrium configuration is the one with the smallest $H_{c1}$.
In this paper we carry out this calculation in
detail assuming two possible arrangements: dilute vortex lines (DVL)
or dilute vortex-line chains (DVLC).

The Gibbs free-energy minimization  for  the DVL  was carried out by
Sudb\o, Brandt and Huse \cite{sbh} in the London limit with the
goal of studying coexistence of vortex 
states at $H_{c1}$. They find that, over a range of parameter values,
there is one  external field tilt angle for which two
DVL states, differing by the vortex-line orientation with respect to
the c-axis, coexist at $H_{c1}$. In this paper we also investigate how these
results are modified by vortex-line chain states.

To carry out the Gibbs free-energy minimization we proceed along lines
similar to those developed by Sudb\o, Brandt and Huse \cite{sbh}. We obtain 
$H_{c1}$ as a function of the external field tilt angle for parameter
values typical of strong and moderate anisotropy,
both for DVL and DVLC, 
and determine which one of these is the equilibrium one
at $H_{c1}$. For strong anisotropy we predict coexistence between DVL
and DVLC  at a particular external field tilt angle.

We consider a bulk uniaxially anisotropic superconductor placed in a
magnetic field ${\bf H}$, with magnitude $H$ and tilted with respect
to the  c-axis  by an angle $\alpha$. We  assume that the vortex
lines are straight and point in a direction making an angle $\theta$
with the c-axis. 

In the London limit the superconductor is
characterized by the penetration depths $\lambda_{ab}$ and
$\lambda_c$,  for currents parallel to the ab-plane and to the
c-direction, respectively. The free energy per unit length
of a generic arrangement of these vortex lines, $F$, can be written as the
sum of pairwise interactions \cite{rev2}  
\begin{equation}
F= \frac{\Phi^2_0}{8\pi}\sum_{i,j} f({\bf r}_i-{\bf r}_j) \; , 
\label{eq.fre}
\end{equation}
where ${\bf r}_i$ is the i-th line  position vector in the plane
perpendicular to the vortex line direction and $f({\bf r})$ is the
Fourier transform of 
\begin{equation}
f({\bf k})= e^{-2g(\bf k)}\;
\frac{1+\lambda^2_{\theta}k^2}{(1+\lambda^2_{ab}k^2)
(1+\lambda^2_{\theta}k^2_x+\lambda^2_c k^2_y)}\; ,
\label{eq.fk}
\end{equation}
where $\lambda^2_{\theta}=\lambda^2_{ab}
\sin^2{\theta}+\lambda^2_c\cos^2{\theta}$,  
$x$ and $y$ refer to two directions  perpendicular to
one another and to the vortex lines, 
with $x$ coplanar with the c-axis, and $g(\bf k)$
is the vortex core cutoff function \cite{rev2,cdk,kog}.

For  both DVL and DVLC,  $F$  can be cast in the form 
\begin{equation}
F_0(N_v,\theta)=n_vA\varepsilon(\theta) \; ,
\label{eq.f0}
\end{equation}
where $n_v$ is the vortex-line density and $A$ is the sample area
perpendicular to the vortex lines direction.
For the DVL we can neglect interactions between
 lines, so that   $\varepsilon(\theta)=\varepsilon_{sf}(\theta)$, the
vortex-line self-energy ($\varepsilon_{sf}(\theta)=
(\Phi^2_0/8\pi) f(0))$. For DVLC we can 
neglect interchain interactions, so that  
$\varepsilon(\theta)=\varepsilon_{sf}(\theta) +
\varepsilon_{ch}(\theta)$, where $\varepsilon_{ch}(\theta)$ is the
energy of interaction per vortex-line of a chain running along the
$x$-direction
($\varepsilon_{ch}(\theta)=(\Phi^2_0/8\pi)\sum_{n\neq0}f(x=nL_1,y=0)$, 
 $L_1$ being the vortex-chain period).   

In order to calculate $H_{c1}(\alpha)$ we have to minimize the Gibbs free
energy with ${\bf H}$ and the sample volume fixed. The Gibbs free energy
per volume for the above described vortex states is given by
\begin{equation}
G_0(B,\theta; H,\alpha)=\frac{B}{\Phi_0}[\varepsilon(\theta)
-\frac{\Phi_0}{4\pi}H\cos{(\theta-\alpha)}] \; ,
\label{eq.gb0}
\end{equation} 
where $B=\Phi_0 n_v$ is the modulus of the magnetic induction. 

We adopt for $\varepsilon_{\rm sf}(\theta)$ the expression derived by
Sudb\o \ and Brandt \cite{sb}, using an elliptic core cutoff function, 
\begin{equation}
\varepsilon_{\rm sf}(\theta)= \varepsilon_0 \frac{\lambda_{\theta}}{\lambda_c}
[\ln({\kappa}{\gamma}) + \frac{\lambda^2_c \cos^2{\theta}}
{\lambda^2_c \cos^2{\theta}+\lambda^2_{\theta}}\ln{\frac{\gamma^2(\lambda^2_c
+\lambda^2_{\theta})}{2\lambda^2_{\theta}}}] \; ,   
\label{eq.esf}
\end{equation}
where  $\varepsilon_0=(\Phi_0/4\pi\lambda_{ab})^2$,
$\gamma=\lambda_{ab}/\lambda_c$, $\kappa=\lambda_{ab}/\xi_{ab}$,  
$\xi_{ab}$ being the ab-plane coherence length.
This line energy  was derived from the anisotropic Ginzburg-Landau theory using the Klemm-Clem transformations \cite{klemm1,klemm}. The elliptical core cut off has semimajor axis $\xi_{ab}^{-1}$, and semiminor axis $(\xi_{ab}^{2}\cos^2\theta + \xi_{c}^{2}\sin^2\theta)^{-1/2}$.

Taking the derivatives of $G_0$ with respect to $B$ and $\theta$ 
and   equating to zero we obtain, respectively 
\begin{equation}
\varepsilon(\theta)-\frac{\Phi_0}{4\pi}H\cos{(\theta-\alpha)}= 0  
\label{eq.ivm1}
\end{equation}
\begin{equation}
\frac{{\rm d}\varepsilon(\theta)}{{\rm
d}\theta}+\frac{\Phi_0}{4\pi}H\sin{(\theta-\alpha)}  =  0 \; .
\label{eq.ivm2}
\end{equation}
The first equation is the familiar condition $G_0=0$ at $H_{c1}$, 
whereas the second results from torque balance.
 
To  solve this system of equations we first obtain $H$ from Eq.\ (\ref
{eq.ivm1})    and substitute it in Eq.\ (\ref {eq.ivm2}). The following
relationship between $\alpha$  and $\theta$ results 
\begin{equation}
\alpha= \theta + \tan^{-1}{[\frac{1}{\varepsilon(\theta)}}\frac{{\rm
d}\varepsilon(\theta)}{{\rm d}\theta}] \; .
\label{eq.alte}
\end{equation}
Finally, we obtain $H_{c1}(\alpha)$  by  substituting the function 
$\theta(\alpha)$  obtained by inverting Eq.\ (\ref {eq.alte})  
into Eq.\ (\ref {eq.ivm1})
\begin{equation}
H_{c1}(\alpha)=\frac{4\pi \varepsilon[\theta(\alpha)]}
{\Phi_0\cos{[\theta(\alpha)-\alpha]}} \; .
\label{eq.hc1al}
\end{equation}
As we shall see shortly, $\theta(\alpha)$ may  be a multivalued
function.
Multivalued solutions for $H_{c1}$ in anisotropic materials have been first noticed 
by Klemm and Clem\cite{klemm1}.
 If this is so, according to Eq.\ (\ref {eq.hc1al}), for each
$\alpha$ there  are several, generally distinct, $H_{c1}(\alpha)$ and
corresponding vortex configurations satisfying the minimization
conditions. The physical solution is the one with the smallest
$H_{c1}(\alpha)$. 

Now we obtain numerically the solutions of these equations for the
following choices of parameters, typical of strong and moderate anisotropy:

\noindent Moderate anisotropy: $\kappa =50$, $\gamma=1/5$

\noindent Strong anisotropy: $\kappa=10$, $\gamma=1/\surd{200}$

\begin{figure}
\epsfxsize=5.0in
\epsfysize=8.0in
\epsfbox{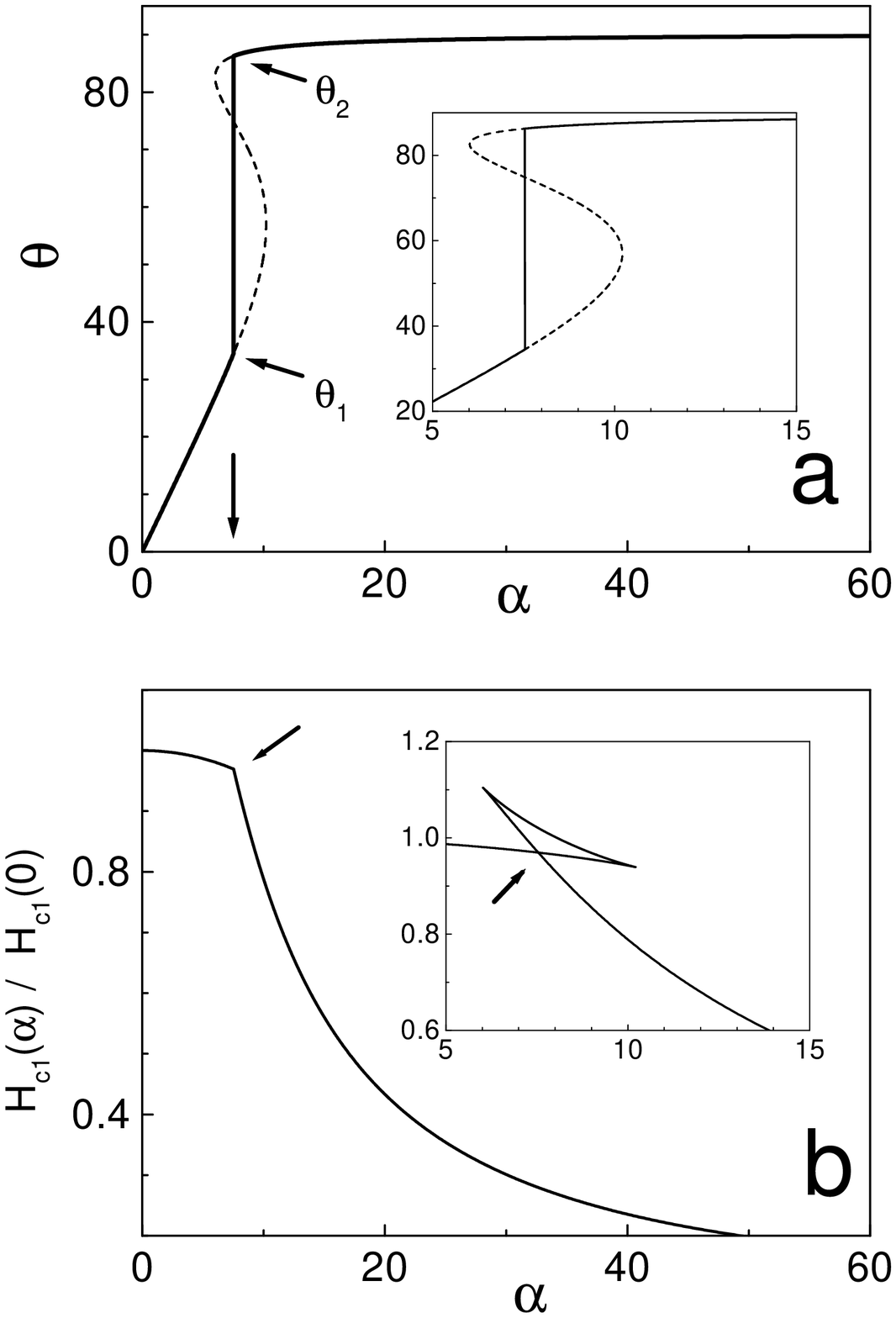}
\caption{ Dilute vortex-line arrangement for strong anisotropy. a)
$\theta(\alpha)$ curve. Dashed line: solution of Eq.\ (\ref {eq.alte}).
Arrow indicates coexistence point ($\alpha=7.5^o$).
Full line: physical solution with smallest $H_{c1}$. b) Smallest
$H_{c1}(\alpha)$ curve. Insets: detail of region where
$\theta(\alpha)$ and $H_{c1}(\alpha)$ are multivalued. Arrows in b)
indicate point where $H_{c1}(\alpha)$ curve changes branches.}    
\label{fig.thabc}
\end{figure}

\begin{figure}
\epsfxsize=5.0in
\epsfysize=8.0in
\epsfbox{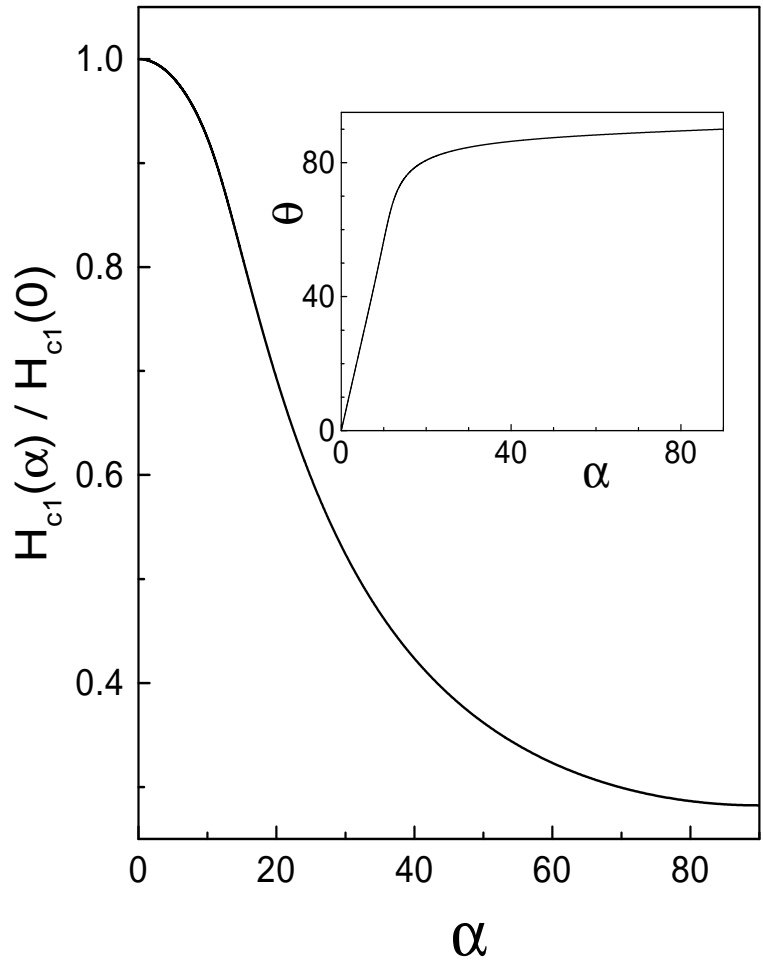}
\caption{ Dilute vortex-line arrangement for moderate anisotropy.} 
\label{fig.thayb}
\end{figure}

\begin{figure}
\epsfxsize=5.0in
\epsfysize=8.0in
\epsfbox{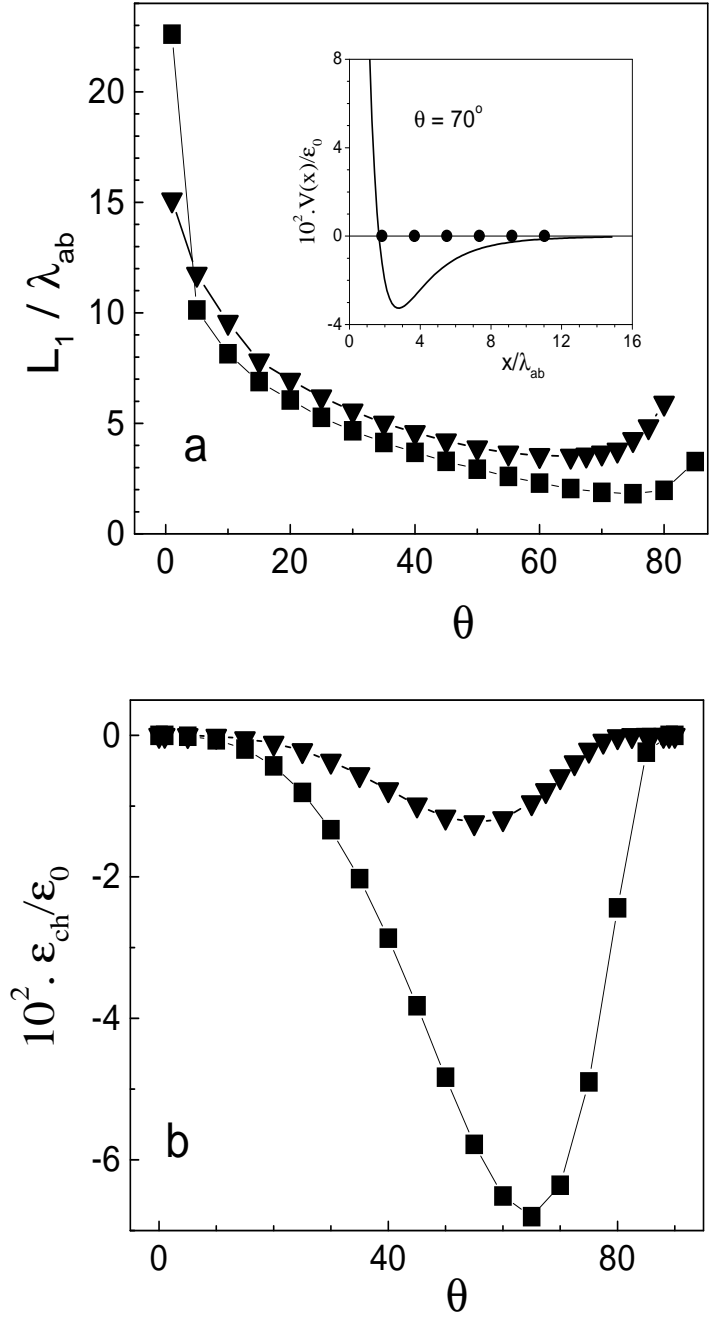}
\caption{\vspace{0.2in} Numerical results for chains. Triangles: moderate anisotropy.
Squares: strong anisotropy. a) Chain period. Inset: interaction
energy of a vortex-line pair along the x-direction (full line) and chain
nearest neighbors positions (full circles). b) 
Chain interaction energy per vortex line.}    
\label{fig.chain}
\end{figure}

\noindent The results are as follows. 

i) {\it Dilute arrangement of vortex lines} (DVL).  As expected, our results
are identical to those obtained by Sudb\o, Brandt and
Huse \cite{sbh}. For strong anisotropy there is  a region 
where $\theta(\alpha)$ is multivalued: for each $\alpha$ there are
three values of $\theta$ and of $H_{c1}(\alpha)$ as shown in Fig.\
\ref{fig.thabc}. The physical solution for each $\alpha$ is the one
with the smallest $H_{c1}(\alpha)$ (Fig.\ \ref{fig.thabc}).  
At $\alpha= 7.5^o$ the solutions $\theta_1=35^o$ and
$\theta_2=86^o$ give the same $H_{c1}$ ((Fig.\ \ref{fig.thabc}),
indicating  coexistence of  DVL with  vortex lines
oriented in  these directions. The region $\theta_1 < \theta <
\theta_2$  is forbidden for vortex-line tilt angles.
For moderate anisotropy there is a single $\theta$ and $H_{c1}(\alpha)$
for each $\alpha$ as   show in Fig.\ \ref{fig.thayb}.   

ii){\it  Dilute arrangement of  vortex-line chains} (DVLC). First we calculate 
$\varepsilon_{\rm ch}(\theta)$   by numerically minimizing 
the energy per vortex line of a single chain with respect to the chain
period, $L_1$,  using a simulated annealing algorithm 
based on a fast convergent series for the energy \cite{wg,mmd}.  The results
for $\varepsilon_{\rm  ch}(\theta)$ and for $L_1$  are
shown in Fig.\ \ref{fig.chain}. Our results are similar to those
obtained in Ref.\cite{vatr1}. 

It turns out  that  $\varepsilon_{\rm ch}(\theta)$ is small compared to
$\varepsilon_{\rm sf}(\theta)$ for all $\theta$: less than 6\% for
strong anisotropy and less than 0.5\% for moderate anisotropy. However,
as we shall discuss in detail shortly, this small correction has
non-trivial  consequences for strong anisotropy.

It is interesting to analyze in some detail the chain structure.
In Fig.\ \ref{fig.chain} we show the interaction energy of a
vortex-line pair and  the  equilibrium positions of the six nearest
neighbors of a given vortex line for strong anisotropy and $\theta
=70^o$. We note that the chain period is considerably smaller than the
distance where the vortex-line pair interaction is minimum. This
results from the long range of the attractive interaction between vortex 
lines.  Indeed, in order to reproduce our calculated value for
$\varepsilon_{ch}$ in this case it is necessary to add the
contributions of all six neighbors shown in Fig.\ \ref{fig.chain}.

We accurately fit our numerical data for $\varepsilon_{\rm 
ch}(\theta)$ to an analytical function
and obtain the solutions of Eqs.\ (\ref {eq.ivm1}) and (\ref {eq.ivm2})
with  $\varepsilon(\theta)=\varepsilon_{sf}(\theta) +
\varepsilon_{ch}(\theta)$ following the same steps as in i).

For moderate anisotropy we find that the $\theta(\alpha)$ and
$H_{c1}(\alpha)$ for the  DVLC   are 
practically identical with the same quantities for the DVL. This
results from the very small $\varepsilon_{\rm ch}(\theta)$ in this case.  

\begin{figure}
\epsfxsize=5.0in
\epsfysize=8.0in
\epsfbox{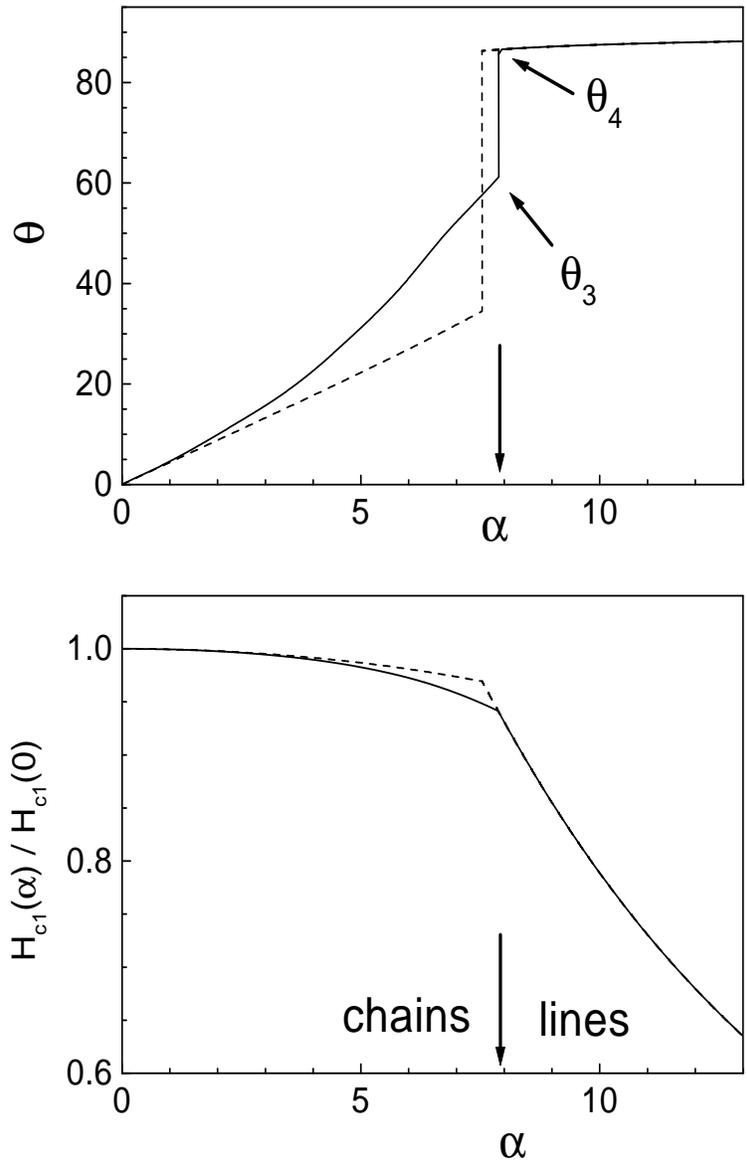}
\caption{Dilute vortex-line chain and vortex-line arrangements for
strong anisotropy compared. Dashed lines:  (smallest value) dilute vortex-line
arrangement (same as in Fig.\ \ref{fig.thabc}. Full line: $\theta(\alpha)$ and
$H_{c1}(\alpha)$ (smallest value) for  dilute vortex-line chain
arrangement. Arrow indicates coexistence point ($\alpha =7.9^o$).}  
\label{fig.thach}
\end{figure}

For strong anisotropy,  
we find that $\theta(\alpha)$ for the DVLC 
is multivalued in a range of $\alpha$ values, with three values of
$\theta$ and $H_{c1}(\alpha)$ for each $\alpha$, similarly  to the DVL case. 
The smallest $H_{c1}(\alpha)$, and the corresponding $\theta(\alpha)$,
are  the full curves shown in Fig.\ \ref{fig.thach}. Comparing the
results for the two  
configurations we conclude that for $\alpha <7.9^o$   
$H_{c1}(\alpha)$ for DVLC is smaller than that for lines, whereas for
$\alpha > 7.9^o$ there is practically no difference between the
$H_{c1}(\alpha)$ for these configurations (the same is true for
$\alpha\sim 0^o$). 
At $\alpha = 7.9^o$ there is coexistence between a DVLC arrangement
with $\theta_3=61^o$ and a DVL one with $\theta_4=\theta_1=86^o$. 
The region $\theta_3 < \theta < \theta_4$  is forbidden for vortex-line 
tilt angles.
  
In conclusion then, we determine if a dilute arrangement of
vortex-lines or of vortex-line chains is the equilibrium vortex configuration
at lower critical field by  calculating $H_{c1}$ for each case and by identifying 
the equilibrium configuration with that for which  $H_{c1}$ is the smallest.
We find that the vortex-line chain configuration is favored for strong
anisotropy and small external field tilt angles ($<7.9^o$). For larger
tilt angles the vortex lines are nearly parallel to the a-b plane, where
there is little difference between the two configurations. For moderate
anisotropy we find no significant difference between the two $H_{c1}$.
The values chosen in our calculation of the anisotropy parameter
$\gamma$   for what we call  moderate and strong anisotropy are 
typical of YBCO and BSCCO, respectively.  The parameter $\kappa$ for
moderate anisotropy is typical of YBCO, but for strong anisotropy is
about five times smaller than those typical of BSCCO \cite{rev1}. This difference
does not alter significantly the above stated conclusions because the
chain interaction energy does not depend on $\kappa$. Only the
self-energy does.

\acknowledgments
This work was supported in part by MCT/CNPq, FAPERJ
and CAPES.

\pagebreak
 
\end{document}